\documentclass{nature_fig}

\usepackage[english]{babel}
\usepackage{bbold}
\usepackage{bbm}
\usepackage[pdftex]{graphicx}
\usepackage{latexsym,amsmath,verbatim}
\usepackage{color}
\usepackage{rotating}
\usepackage{multirow}
\usepackage{color}
\usepackage{caption}

\usepackage{amssymb,amsfonts,amsmath}
\renewcommand{\(}{\left(}
\renewcommand{\)}{\right)}

\newcommand{\tr}[1]{\text{Tr}\(#1\)}
\renewcommand{\(}{\left(}
\renewcommand{\)}{\right)}
\newcommand{\heading}[1]{{\vspace{0.25truecm}\noindent\textbf{#1.}}}
\usepackage{xcolor}
\definecolor{RoyalBlue}{HTML}{4169e1}
\definecolor{ForestGreen}{HTML}{228b22}

\usepackage[colorlinks = true,
            linkcolor = blue,
            urlcolor  = blue,
            citecolor = blue,
            anchorcolor = blue]{hyperref}

\usepackage{comment}
\usepackage{subcaption}

\usepackage{todonotes}

\definecolor{mypink2}{RGB}{219, 48, 122}

\title{Unraveling the effects of multiscale network entanglement on disintegration of empirical systems}

\author{Arsham Ghavasieh$^{1}$, Massimo Stella$^{1,3}$ , Jacob Biamonte$^{2}$, Manlio De Domenico$^{1\ast}$}

\begin{document}

\maketitle

\begin{affiliations}
\item CoMuNe Lab, Fondazione Bruno Kessler, Via Sommarive 18, 38123 Povo, Italy
\item Skolkovo Institute of Science and Technology, 3 Nobel Street, Moscow 121205, Russia
\end{affiliations}
\vspace{-0.5truecm}
\noindent Corresponding author: mdedomenico@fbk.eu

\footnotetext[3]{Now at Complex Science Consulting, Via Amilcare Foscarini 2, 73100 Lecce, Italy}

\baselineskip24pt

%http://www.nature.com/nphys/authors/article_types/index.html

\begin{abstract}
Complex systems are large collections of entities that organize themselves into non-trivial structures that can be represented by networks. A key emergent property of such systems is robustness against random failures or targeted attacks ---i.e. the capacity of a network to maintain its integrity under removal of nodes or links. Here, we introduce network entanglement to study network robustness through a multi-scale lens, encoded by the time required to diffuse information through the system. Our measure's foundation lies upon a recently proposed framework, manifestly inspired by quantum statistical physics, where networks are interpreted as collections of entangled units and can be characterized by Gibbsian-like density matrices. We show that at the smallest temporal scales entanglement reduces to node degree, whereas at the large scale we show its ability to measure the role played by each node in network integrity. At the meso-scale, entanglement incorporates information beyond the structure, such as system's transport properties. As an application, we show that network dismantling of empirical social, biological and transportation systems unveils the existence of a optimal temporal scale driving the network to disintegration. Our results open the door for novel multi-scale analysis of network contraction process and its impact on dynamical processes.
\end{abstract}

%\section*{\rev{Main}}

%%%%%%%%%%%%%%%%%%%%%%%
%Short intro or abstract follow-up
%%%%%%%%%%%%%%%%%%%%%%%

%%%%%%%%%%%%%%%%%%%%%%%
% Corpus
%%%%%%%%%%%%%%%%%%%%%%%

%%%%%%%%%%%%%%%%%%%%%%%
%Conclusions
%%%%%%%%%%%%%%%%%%%%%%%

%%%%%%%%%%%%%%%%%%%%%%%
%%%%%%%%%%%%%%%%%%%%%%%
%%%%%%%%%%%%%%%%%%%%%%%
A key characteristic of complex systems, as large collections of interconnected entities, is their robustness against damage, whether it is genetic mutations in gene-gene interaction networks\cite{manke2006entropic}, extinction of species in ecosystems\cite{dunne2002network}, failure of internet routers\cite{doyle2005robust} or unavailability of transportation means\cite{de2014navigability}. This common property might be deeply rooted in the unexpected resistance of their structures to disintegration \cite{albert2000error,callaway2000network,holme2002attack,centola2008failure,trajanovski2013robustness}. The structure is often represented by networks, where nodes play the role of entities and links specify the connections between them. In this framework, network integrity---defined as the availability of link sequences connecting every pair of nodes--- and its maintenance has proven fundamental for correct functioning of the system as a whole\cite{iyer2013attack,de2014navigability}. 

In contrast, disintegration is often caused by internal failures or external attacks, widely, modeled in terms of progressive removal of nodes or links. Consequently, a network contracts and dismantles into a number of components of different sizes, each containing a number of interconnected nodes while being disconnected from other components. As the shrinking process proceeds, the size of the largest of these connected component (LCC) decays until it vanishes and the network dismantles into isolated nodes.

The size of LCC has been widely adapted as a proxy for network robustness under random or targeted removals\cite{albert2000error,holme2002attack,iyer2013attack}. The latter includes identifying the central nodes and detaching them according to their ranking, aiming for the maximum possible damage. Asking which set of node is relevant for a fast disintegration led to various definitions and proxies where none outperforms the others in every scenario--- e.g. ranking based on the betweenness centrality proves more effective for certain classes of networks\cite{holme2002attack,estrada2006network}. It is however surpassed by the degree centrality in other cases\cite{iyer2013attack}. More recent sophisticated descriptors are available and often work well in a range of scenarios~\cite{morone2015influence,braunstein2016network,ren2019generalized}. A systematic evaluation of state-of-the-art methods reveals that the best approach, based on iterative betweenness, is one of the oldest ones but also one of the most computationally expensive, making it unsuitable for large networks~\cite{wandelt2018comparative}.
The problem of optimal percolation and network dismantling  remains open, while the aim of this article is to provide a novel framework to study the disintegration process emphasizing two points, seemly missing in the literature. First, most centrality measures rely on network descriptors such as degree or shortest path. Evidently, the information content of a network as a whole can not be fully captured by these proxies. Second, the importance of network integrity, and maintaining it under damage, is to sustain the node-node communications. Thus, understanding the information exchange among the nodes beyond shortest-path communication, and how it is affected in the disintegration process, requires a multi-scale framework---e.g. to differentiate between the short- and long-range signalling between the nodes, not necessarily passing through the shortest paths, as captured by the betweenness descriptor. Therefore, the main research question of our work is not limited to defining a novel centrality measure and compare its impact on network robustness. Instead, we are interested in better understanding if operators such as the network density matrix, inspired by quantum statistical physics and information theory~\cite{de2016spectral,biamonte2019complex,ATransport2020}, are able to capture the main features of communication flows, beyond shortest paths, and exploit them to better characterize system's resilience to targeted attacks.  

To this aim, we propose network entanglement, described by a Gibbsian-like density matrix~\cite{de2016spectral} which is derived from the propagator of diffusion dynamics, with a tunable parameter $\beta$ encoding the propagation time and playing the role of a multi-scale lens. In the following we show the properties of our measure at the micro-, meso- and macro-scale, while demonstrating the existence of an information-theoretic optimal scale, $\beta_{c}$, at which node's impact is determined by its role in the transport properties of system. At this scale, we study the disintegration of a range of synthetic networks as well as real-world social, biological and transportation networks, to show that dismantling is always comparable  with the one obtained from other approaches across different scenarios. 

\section*{Results}
\textbf{Theoretical grounds.} The information content of complex networks can not be fully captured by means of traditional descriptors such as the degree distribution and diameter. For this reason, a variety of tools and methods have been introduced with roots in statistical physics and information theory\cite{Cimini2019}.

Recently, it has been shown that networks can be viewed as collections of entangled entities represented by a grounded density matrix resembling the Gibbs state\cite{de2016spectral,Biamonte2019} that is used, successfully, to analyse a range of empirical networks from transportation systems\cite{ATransport2020} to the human microbiome\cite{de2016spectral} and brain\cite{Nicolini2020}. 

The Gibbsian density matrix of a network $G$ with $N$ nodes, represented by an adjacency matrix $A$ ($A_{ij}=1$ if nodes $i$ and $j$ are connected, it is 0 otherwise ), has been originally proposed\cite{de2016spectral} as the exponential function of the combinatorial Laplacian matrix ${L}=D-A$, where $D$ is a diagonal matrix defined by $D_{ii}=k_{i}$ and $k_{i}=\sum_{j}A_{ij}$ denotes the degree of $i-$th node, as follows
\begin{eqnarray}\label{eq:density}
\rho_\beta = \frac{e^{-\beta { L}}}{\tr{e^{-\beta { L}}}},
\end{eqnarray}
in terms of the ratio between the propagator of diffusion dynamics on top of the network, with $\beta$ encoding the diffusion time, and its trace encoding the partition function $Z_{\beta}=\tr{e^{-\beta { L}}}$, which plays an important role in the transport properties of networks~\cite{ATransport2020}. Using Eq.~\ref{eq:density} the Von Neumann entropy can be obtained as 
\begin{eqnarray}\label{eq:entropy}
S_{\beta}(G) = -\tr{\rho_\beta\log_{2}\rho_\beta}.
\end{eqnarray}

Recently, a mean-field approximation of the Von Neumann entropy has been introduced to simplify the many term summation and allow for analytical derivations\cite{ATransport2020}. However, that approximation is limited to the case of random walk dynamics and can not be used for the purpose of this article. Consequently, here, we derive a mean-field entropy (See Methods) that is valid for the case of continuous diffusion: 
\begin{eqnarray}\label{eq:MF_precise}
S^{MF}_{\beta} = \frac{1}{\log{2}} ( \beta \frac{2m}{N-C} \frac{Z_{\beta}-C}{Z_{\beta}} + \log Z_{\beta} ). 
\end{eqnarray}
where $C$ is the number of disconnected components of the network and $m$ is the overall number of links. In most networks, the number of nodes is much larger than the number of disconnected components $N\gg C$, and, therefore, Eq.~\ref{eq:MF_precise} can be approximated as
\begin{eqnarray}\label{eq:MF}
S^{MF}_{\beta} = \frac{1}{\log 2} ( \beta \bar{k} \frac{Z_{\beta}-C}{Z_{\beta}} + \log Z_{\beta} ), 
\end{eqnarray}
where $\bar{k}$ is the mean degree of nodes. Also, in case of large $\beta$, the mean field entropy reduces to (See Methods) 
\begin{eqnarray}\label{eq:MF_approx}
S^{MF}_{\beta} \approx ( \beta \bar{k} + 1 ) \log_{2}{Z_{\beta}}. 
\end{eqnarray}

\begin{figure}
 %\captionsetup{width=0.5\textwidth}
 \centering
 \includegraphics[width=0.75\linewidth]{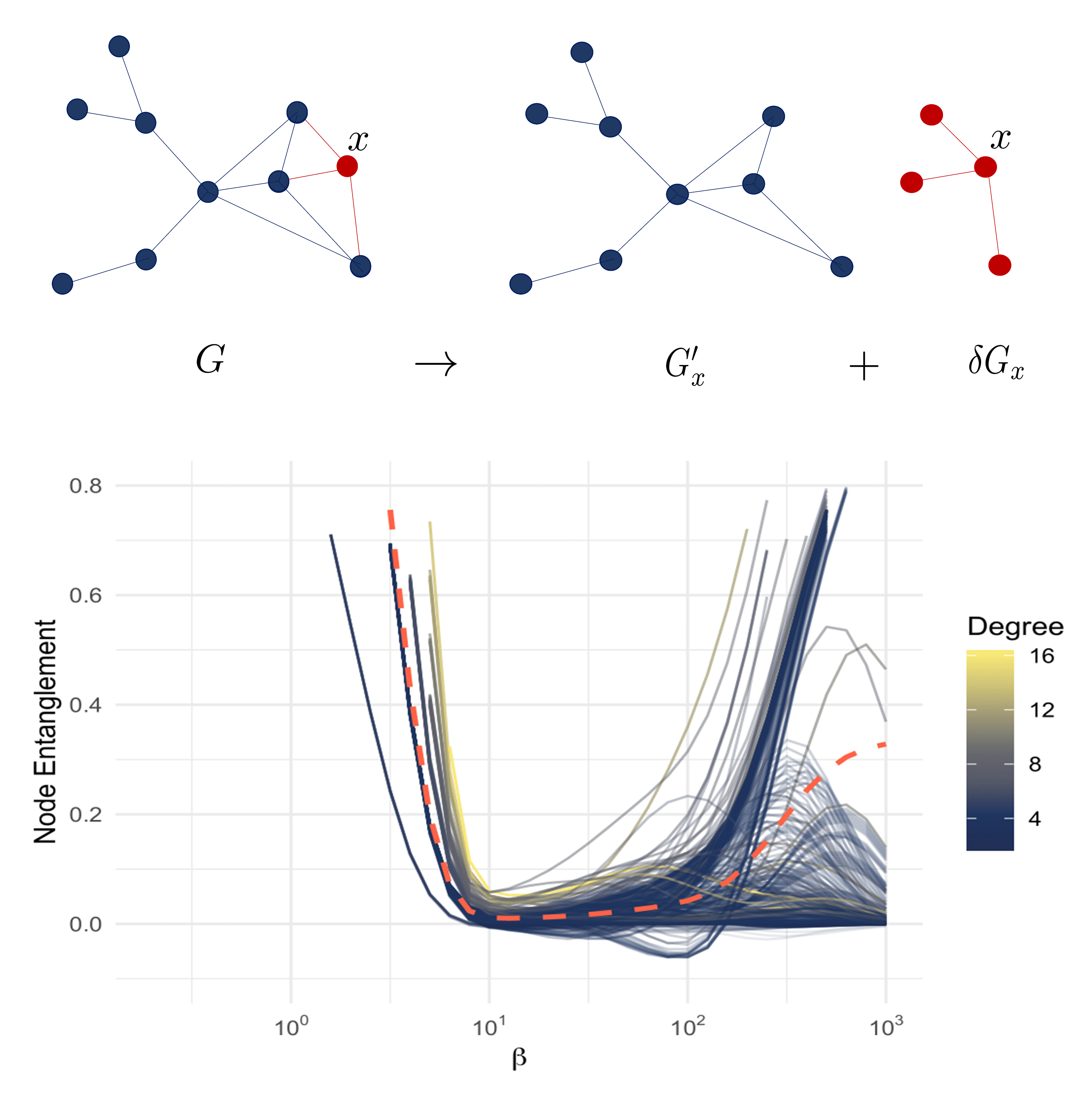}
 \caption{\textbf{Detachment process.} The process of detaching node $x$ and its incidence edges from the original network $G$ is plotted (top). The entanglement of each node is shown as a function of Markov time $\beta$, for an arbitrary network (bottom). Each trajectory is colored according to the degree of the detached node, to highlight that there is no trivial relationship between entanglement and degree across scale. The collective entanglement $\bar{M}_{\beta}$, defined as average entanglement of nodes, is shown by orange dashes.\label{fig:disintegration}}
\end{figure}

\textbf{Defining network entanglement.} To quantify the importance of a single node $x$ in the interconnected system, we, firstly, detach it from the network $G$ with its corresponding incident edges. The removed node and its incident edges form a star network, indicated by $\delta G_{x}$, having the size $k_{x} + 1$ where $k_{x}$ is the degree of node $x$. The remainder of $G$ shapes the perturbed network $G'_{x}$, that has $N-1$ nodes (See Fig.~\ref{fig:disintegration}).

We define the {\it entanglement} between each node $x$ and the network as 
\begin{eqnarray}\label{eqn:entanglement}
M_{\beta}(x) = [S_{\beta}(G'_{x}) + S_{\beta}(\delta G_{x})] - S_{\beta}(G),
\end{eqnarray}
By tuning the propagation time $\beta$, the entanglement between the nodes and network is expected to change. Using Eq.~\ref{eq:entropy} and ~\ref{eq:MF_approx}, we show (See Methods) that in extreme cases, the entanglement centrality follows:
\begin{itemize}
    \item $\beta \rightarrow 0 :$ $M_{\beta}(x) \approx \log_{2}{(k_{x}+1)}$
    \item $\beta \rightarrow \infty :$ $M_{\beta}(x) \approx \beta \bar{k} \log_{2}{C'_{x}}  $
\end{itemize}

where $k_{x}$ is the degree of the removed node and $C'_{x}$ is the number of disconnected components, in the perturbed network $G'_{x}$. Clearly, if the entanglement is used as a centrality measure, it coincides with the degree centrality at small scales. It is worth remarking here that a network has highest integrity if it has only one connected component---i.e. for every pair of nodes, there is at least one link or sequence of links (path) that connects them. Therefore, at the large scale,  entanglement centrality evaluates the direct role of nodes in keeping the integrity of network, by considering the number of disconnected components generated consequent to their detachment.

The intermediate scales exhibit even richer information. To better characterize this information, we define the collective entanglement as the average entanglement of all the nodes $\bar{M}_{\beta} = \frac{1}{N}\sum\limits_{x=1}^{N}M_{\beta}(x)$ (See Fig.~\ref{fig:disintegration}). 

Let us assume that this collective variable reaches its minimum at some optimal scale $\beta_{c}$, which is still unknown. We analytically show (See Methods) that the centrality of any node $x$, near $\beta_{c}$, is proportional to the change in the partition function $Z_{\beta}$ caused by its detachment:
\begin{equation} 
M_{\beta_c}(x) \approx \frac{\beta_c \bar{k} + 1 }{N Z_{\beta_c} \log{2}} \Delta Z_{\beta_c}(x),
\end{equation}
where $\Delta Z_{\beta}(x) = Z'_{\beta}(x)-Z_{\beta}$, and $Z'_{\beta}(x)$ is the partition function of the perturbed networks $\Delta G'_{x}$. The partition function $Z_{\beta}$ has been recently related to dynamical trapping of information flow within a system topology, to assess the transport properties of complex networks~\cite{ATransport2020}. Therefore, at this scale, a node is more central if its removal hinders the diffusion within the network more effectively than other ones.

In the following we study the dismantling process at the temporal scale $\beta_{c}$. Yet, it is worth mentioning that entanglement centrality provides a meaningful measure in other choices of $\beta$ which are discussed so far.

\begin{figure}
 \centering
 \includegraphics[width=\linewidth]{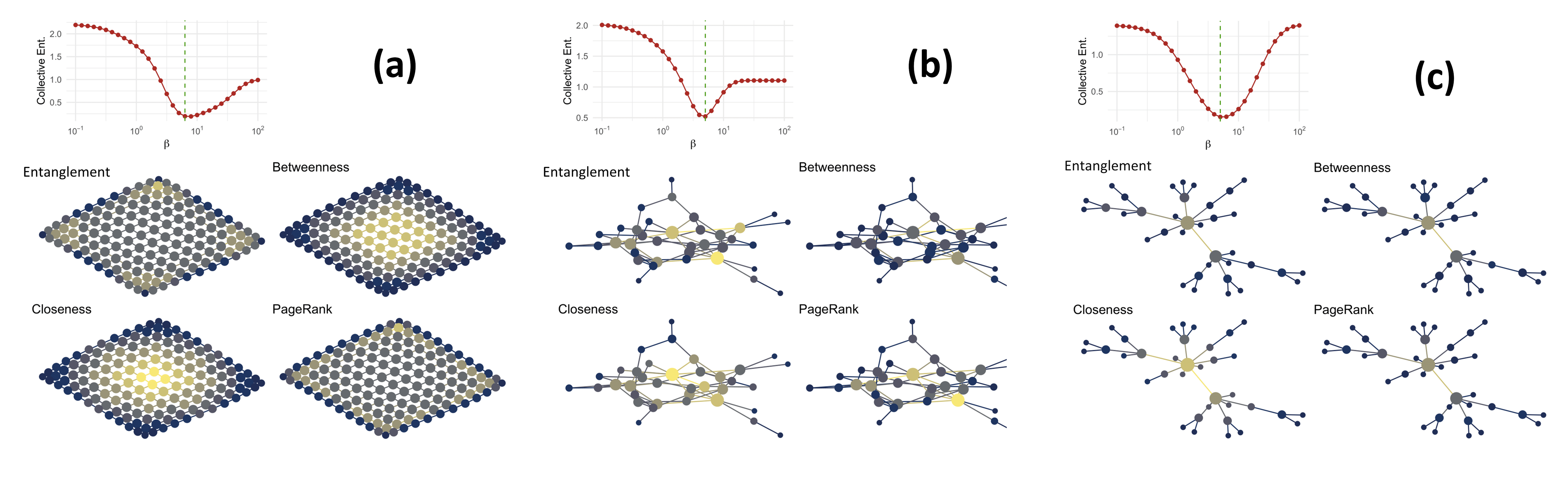}
\caption{\textbf{Entanglement as a multi-scale measure of node centrality.} A lattice (a), an Erdos Renyi (b) and a Barabasi-Albert network (c) are considered, from left to right panels. The centrality of each node is color coded from lighter to darker according to distinct measures:  entanglement, betweenness, closeness and  PageRank. The specific time scale $\beta_{c}$ has been considered in case of entanglement centrality, by minimizing the collective measure $\bar{M}_{\beta}$, here shown in the top panels. It is evident that network entanglement is not trivially related to existing centrality measures.\label{fig:comparison}}
\end{figure}

\begin{figure}
\centering
\includegraphics[width=\linewidth]{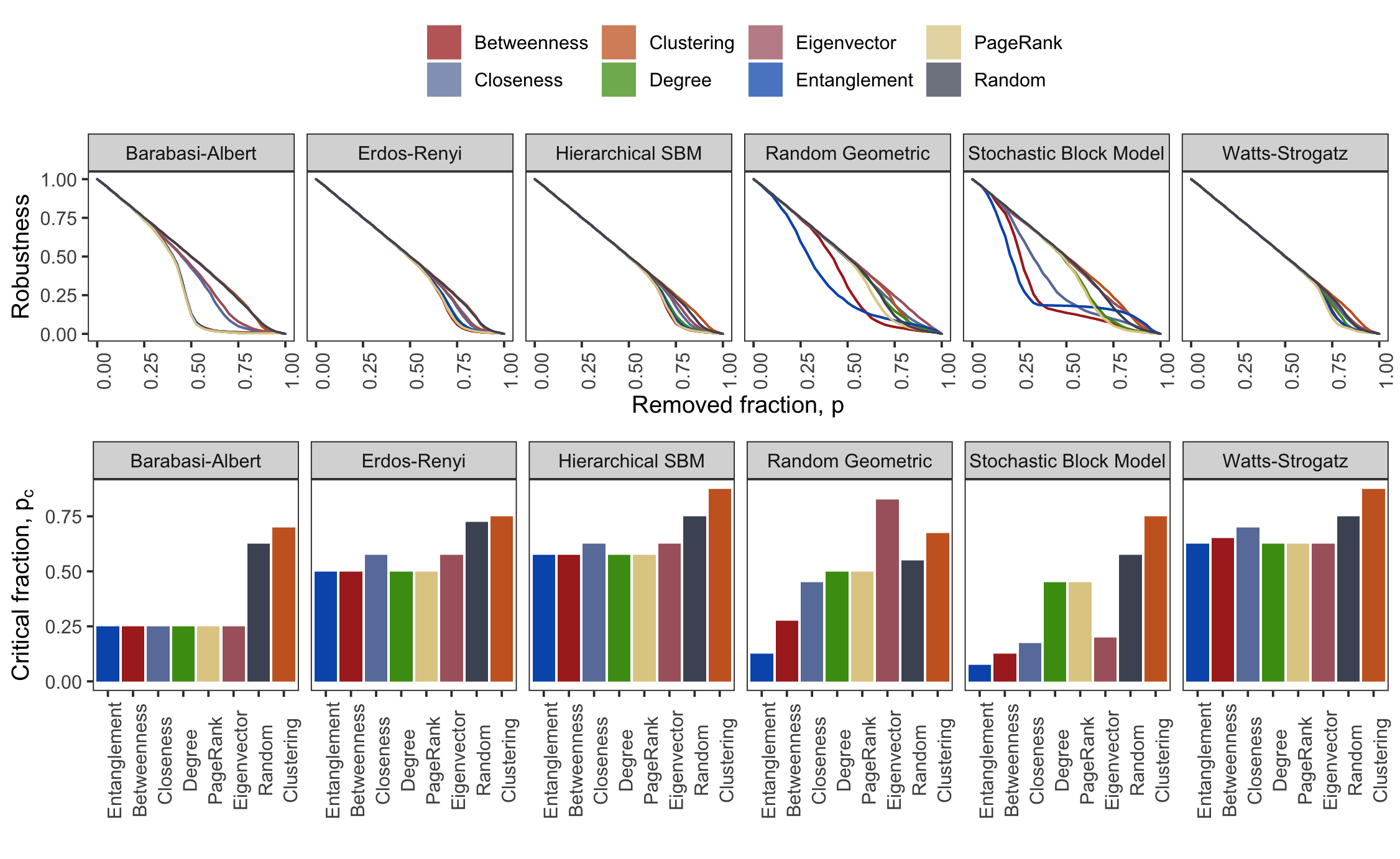}
\caption{ \textbf{Entanglement analysis of synthetic networks.} Disintegration of different network topologies, including Barabasi-Albert, Eros-Renyi, hierarchical stochastic block model, random geometric, stochastic block model and Watts-Strogatz models. The robustness of an ensemble of each network model is tested against random failures and targeted attacks based on seven  measures of node centrality including betweenness, clustering, eigenvector, PageRank, closeness, degree and entanglement, the last one defined in this study. entanglement centrality, tuned at time-scale $\beta_{c}$, performs equal or faster than other measures in breaking the network into its critical fraction, as shown in the bottom boxes, where bars are ordered according to the overall performance ---across all numerical experiments ---of each measure.\label{fig:synthetic}}
\end{figure}

\textbf{Entanglement analysis of synthetic networks.} We consider six different classes of networks, including Barabasi-Albert\cite{Barabsi1999}, Erods-Renyi, hierarchical stochastic block model, random geometric, stochastic block model and Watts-Strogatz\cite{Watts1998} models, frequently used to mimic the topology of natural and man-made complex systems~\cite{boccaletti2006complex}. For each model, an ensemble of 20 independent realizations of $N= 256$ nodes has been considered. We have kept the average degree approximately equal to 12, to allow for a more meaningful comparison across models. The nodes have been ranked according to different measures, including betweenness, clustering, eigenvector, PageRank, closeness, degree (see Methods for details). Finally, we minimized the collective entanglement (See Fig.~\ref{fig:disintegration}) for each network to find its $\beta_{c}$ and used it to find the entanglement centrality for each nodes, according to Eq.~\ref{eqn:entanglement}. The procedure is schematically represented in Fig.~\ref{fig:comparison}.

The results clearly show that the entanglement centrality performs as effective as or faster than the other measures considered here in dismantling the network up to its critical fraction, the point at which the network starts to break into disconnected components (See Fig.~\ref{fig:synthetic}). Remarkably, for random geometric and stochastic block model networks, the disintegration happens significantly faster when using network entanglement. In the case of Barabasi-Albert networks, after the critical fraction, betweenness and entanglement centrality at $\beta_{c}$ act significantly slower than degree and PageRank centrality. As the ranking provided by entanglement centrality at small temporal scales and degree centrality are proven to be identical, this result suggests that the most effective scale for dismantling this class of network, structurally, is smaller than $\beta_{c}$. Nevertheless, the intermediate scale $\beta_{c}$ has been shown effective in all considered cases, outperforming other measures up to the critical fraction.

\textbf{Entanglement analysis of real-world networks.} We investigate the disintegration of a variety of real-world complex networks, representing the structure of biological, transportation and social systems, under progressive attacks based on the discussed centrality measures.

Data sets include the socio-patterns network ($N=241$) representing people attending an exhibition linked by their face to face interactions~\cite{konect:2016:sociopatterns-infectious,konect:sociopatterns,konect}, the Haggle network ($N=274$) representing people and their contacts via wireless devices~\cite{konect,konect:2016:contact_haggle,konect:chaintreau07_haggle}, the New York city transportation network ($N=433$) representing subway stations and their connection~s\cite{Roth2012}, the US airports network ($N=500$) representing the busiest commercial airports in the United States in 2002 with links encoding the flights between them, weighted by the number of seats available at the airplane~\cite{Colizza2007} and the neural network of the nematode worm \textit{C. elegans} ($N=282$) representing neurons linked by their neural junctions~\cite{Watts1998}.

\begin{figure*}
    \begin{subfigure}[t]{0.5\textwidth}
        \centering
        \includegraphics[width=\linewidth]{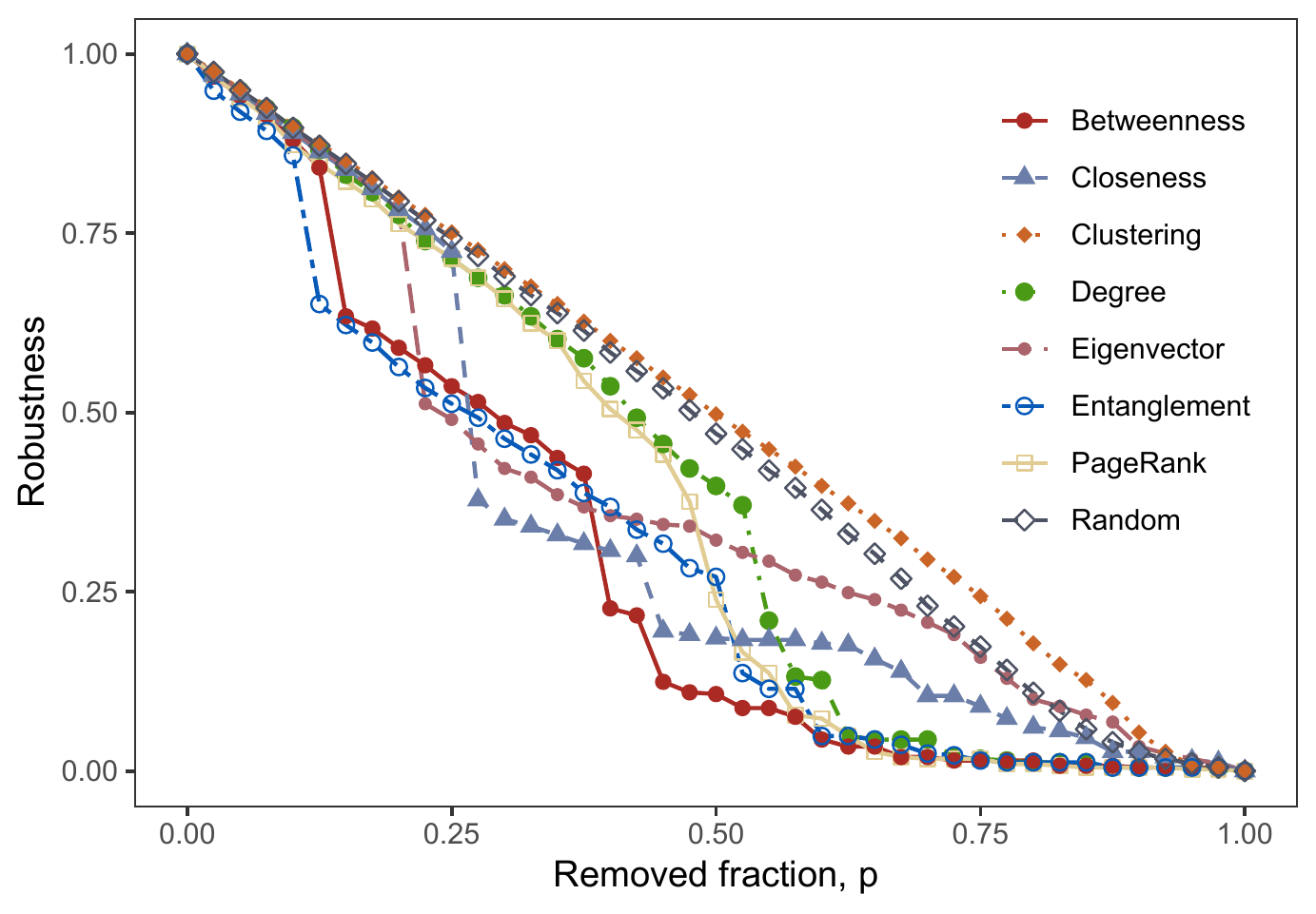}
        \caption{Socio-patterns ($\beta_c=8$)}
    \end{subfigure}
    \begin{subfigure}[t]{0.5\textwidth}
        \centering
        \includegraphics[width=\linewidth]{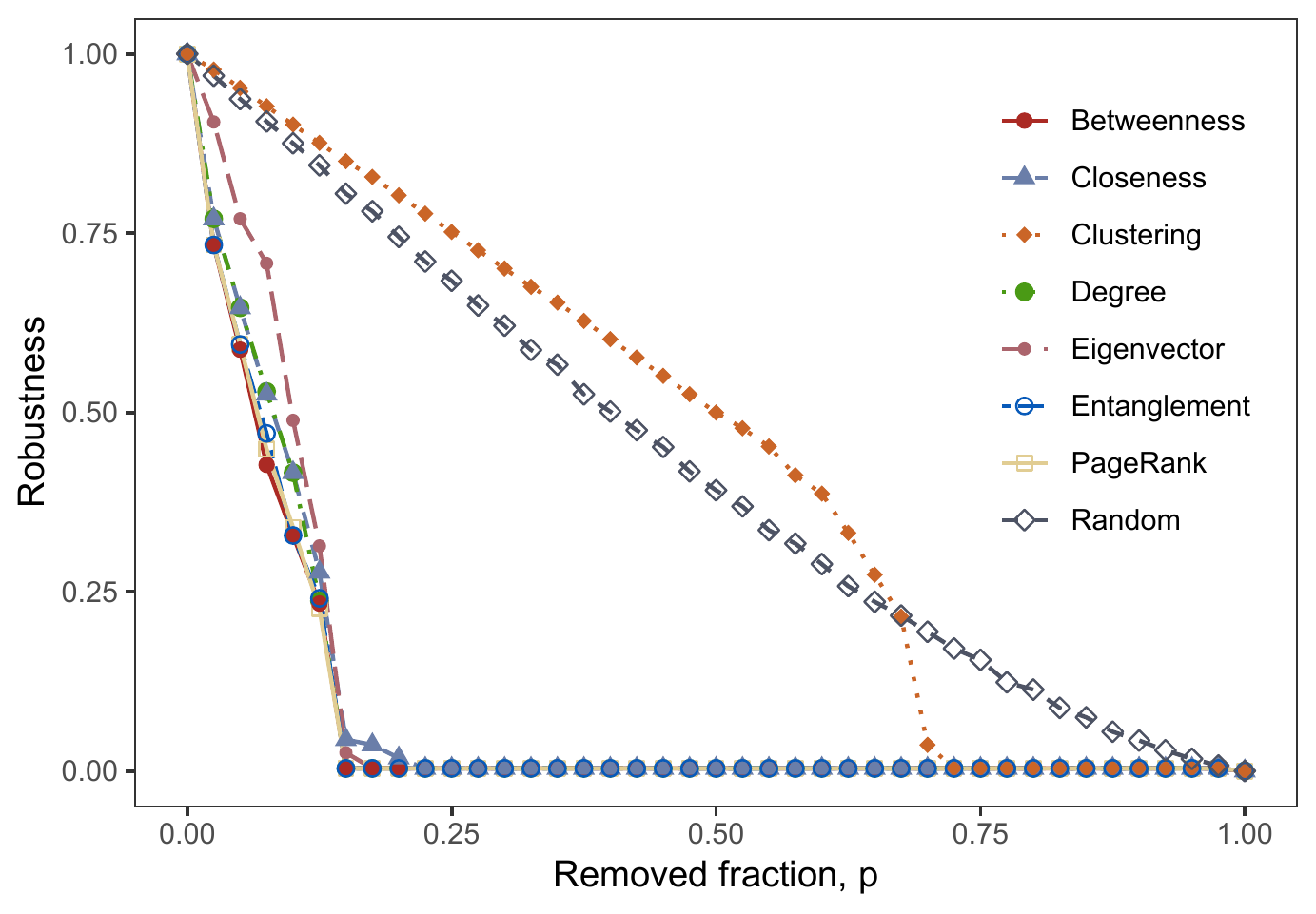}
        \caption{Haggle ($\beta_c=6$)}
    \end{subfigure}
    
    \begin{subfigure}[t]{0.5\textwidth}
        \centering
        \includegraphics[width=\linewidth]{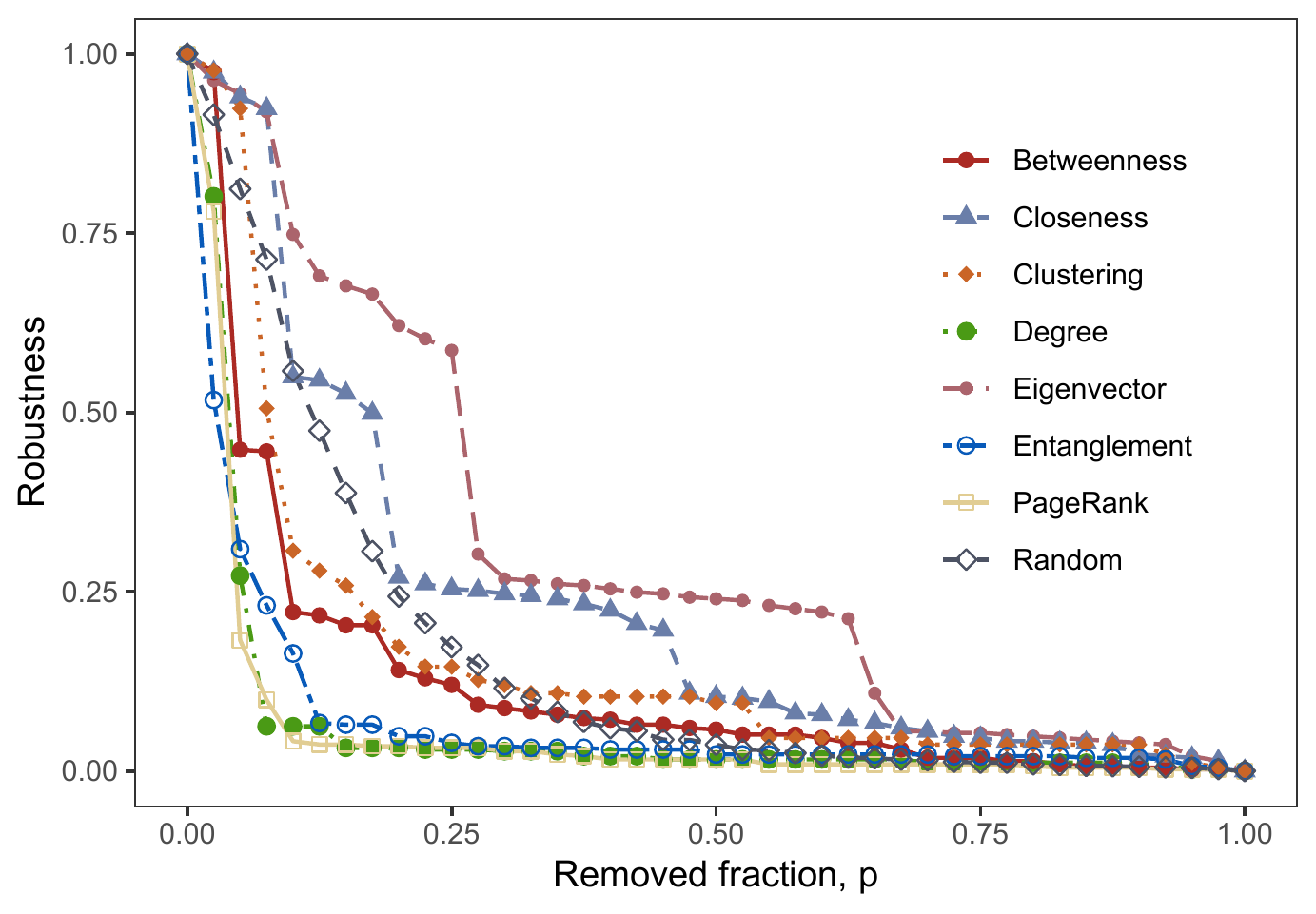}
        \caption{NYC metro ($\beta_c=.4$)}
    \end{subfigure}
    \begin{subfigure}[t]{0.5\textwidth}
        \centering
        \includegraphics[width=\linewidth]{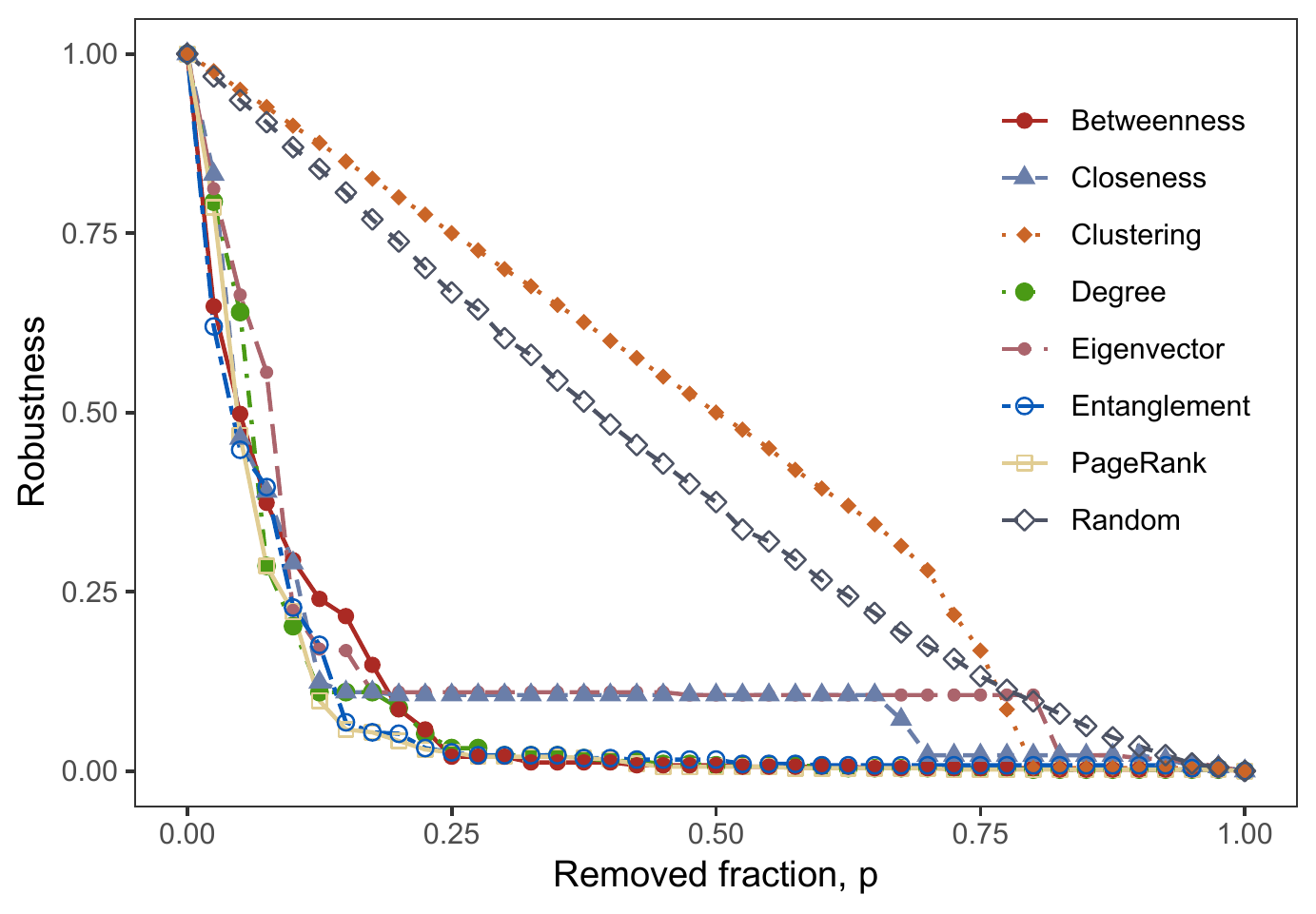}
        \caption{US Airports ($\beta_c=6$)}
    \end{subfigure}
    
    \begin{subfigure}[t]{\textwidth}
        \centering
        \includegraphics[width=.5\linewidth]{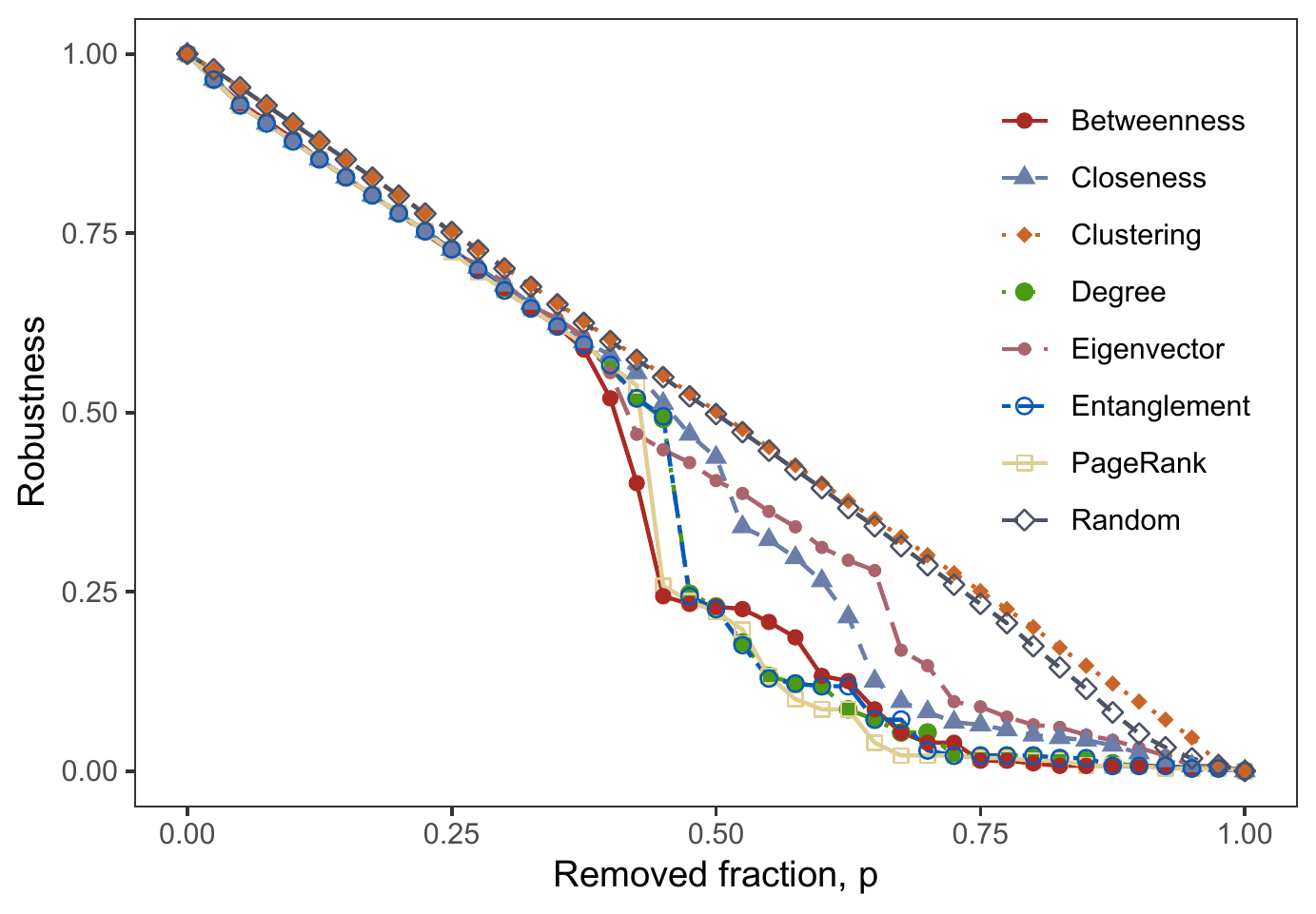}
        \caption{Neural network of C.elegans ($\beta_c=0.9$)}
    \end{subfigure}
\caption{\label{fig:empirical}\textbf{Real-world networks} The robustness of empirical systems under targeted attacks has been pictured. In all cases, the entanglement centrality, tuned at $\beta_{c}$, performs better or equal to all the other discussed measures.}
\end{figure*}

As expected, all these real-world networks show high robustness against random node removals, implying their ability to maintain their function under random failures. However, adopting the right targeted attack strategy can effectively disintegrate them (See Fig.~\ref{fig:empirical}). Although degree and PageRank centrality perform better than other classical measures, in dismantling the transportation networks, such as NYC metro and US airports network, they are outperformed by betweenness and closeness centrality in the system. This result highlights the lack of a universal attack strategy that can be considered always valid, regardless of network features. Interestingly, our analysis indicates that, for all the considered empirical systems, the entanglement centrality provides an effective dismantling strategy (See Fig.~\ref{fig:empirical}), comparable with the best measures and outperforming the others, thus providing a promising candidate for such a universal attack strategy.

\section*{Discussion}

Analyzing the robustness of complex systems is still a challenging task. Here, we have used Gibbsian-like density matrices to quantify the entanglement between nodes and their networks, in order to characterize the impact of node removal on system function. To this aim, we measure the change in the von Neumann entropy of a network, caused by the detachment of nodes and their incident edges, and we have used the nodes' entanglement as a proxy for their centrality in the flow exchange across the network. Our framework is multi-scale, with Markov time $\beta$ playing the role of a tunable parameter which allows one to study the response of the network at micro-, meso- and macro-scales. To this aim, we have developed a mean-field approximation of entropy to analytically explain the behavior of entanglement centrality at different scales. Remarkably, the network entanglement defined in terms of a collective variable allows one to reveal the existence of a characteristic temporal scale $\beta_c$ at which flow exchange can be (sub-)optimally used for efficient network dismantling. 
Our results indicate that for small temporal scales $\beta \rightarrow 0$, the degree of each node determines its entanglement with the network, and entanglement centrality coincides with the well-known degree centrality. At very large scales $\beta \rightarrow \infty$, entanglement centrality measures the direct role of each node in the integrity of network ---i.e. how many disconnected components will appear if the node is detached. Finally, we have shown that the collective entanglement ---i.e. the average entanglement of all the nodes with the network ---reaches its minimum at a specific choice of $\beta = \beta_{c}$. Interestingly, at this scale, we demonstrate that entanglement centrality is rather sensitive to the node's impact on the diffusion dynamics on top of the network, and not the structure. More specifically, according to our measure, a node is ranked higher if its detachment causes a larger increase in the partition function of the system. The partition function provides a proxy for  \emph{dynamical trapping}, an important transport property that indicates the tendency of network to hinder the flow of information~\cite{ATransport2020}: therefore, strategies can be designed to lower the partition function and, consequently, enhance the diffusive flow among nodes. Conversely, here, we target the nodes according to entanglement centrality, aiming for maximum increase in the partition function that, consequently, hinders transport properties.

Of course, the detachment of nodes during the disintegration process alters the topology and, as a consequence, the importance of the remaining nodes. For this reason, adaptive attack strategies ---where the centrality of each node is re-calculated after each perturbation is applied ---become interesting. However effective, they are computationally slow, especially, in case of large networks~\cite{wandelt2018comparative}. Thus, we adopt the static ---i.e., non-adaptive ---attack strategy in this work ---i.e. ranking of nodes according to each centrality measures is calculated only once, at the beginning of disintegration process. Despite this apparent limitation, we show that network entanglement is still able to capture higher-order interactions that are exploited to efficiently dismantle a network.  

The analysis of both synthetic and real-world networks, where different attack strategies are compared to network entanglement at $\beta_{c}$ indicates that our measure performs as well as or faster than other measures, in damaging the network up to its critical fraction, across a range of scenarios. However, it becomes slower than some other measures, after the critical fraction is reached, yet still comparable to the others. This result indicates that entanglement can be used to quickly disrupt the flow exchange, but can not be used to disintegrate a system faster than more traditional approaches.

As mentioned before, the entanglement centrality at $\beta_{c}$ aims to disrupt the dynamics on top of the network, by hindering the diffusive flow. Therefore, a plausible interpretation of our numerical experiments is that disrupting the dynamics comes along with the dismantling of the structure, up to the critical fraction. 

Overall, the presented framework opens the doors for further investigation of the network contraction process, from a multi-scale perspective, and its relation with the dynamics and transport properties of the complex systems.

\section*{Methods}
\textbf{Mean-field entropy.} A mean-field approximation of the network Von Neumann entropy has been recently suggested for the random walk based density matrices~\cite{ATransport2020}. Similarly, here, we derive a mean-field entropy for the case of continuous diffusion. The eigenvalue spectrum of the Laplacian follows:
\begin{itemize}
\item $0 = \lambda_{1} \leq \lambda_{2} \leq ...\leq \lambda_{N}$
\item $\tr{{L}}=\sum\limits_{i=1}^{N} \lambda_{i} = \sum\limits_{i=1}^{N} k_{i} = 2m  $
\end{itemize}
where $m$ is the number of links in the network, where no self loops exist.

At this step, it is worth noting that $\rho_{\beta}$ and ${L}$ can be eigen-decomposed as follows:
\begin{eqnarray}
{L}&=& Q\Lambda Q^{-1},\\
\rho_{\beta} &=&  Q\frac{e^{-\beta\Lambda}}{Z_{\beta}} Q^{-1},
\end{eqnarray}
being the columns of $ Q $ the eigenvectors of the Laplacian matrix and $\Lambda$ is the diagonal matrix of eigenvalues of the Laplacian matrix. For the density matrix, the eigenvalues follow $\nu_{i}(\beta)=e^{-\beta \lambda_{i}}/Z_{\beta}, i=1,2...,N$. The Laplacian matrix and the density matrix can be eigen-decomposed simultaneously, in the basis of eigenvectors of the Laplacian matrix. 

Furthermore, Eq.~\ref{eq:entropy} can be rewritten as:
\begin{eqnarray}
S_{\beta}(G) &=& - \tr{\rho_{\beta}\log_{2}{\rho_{\beta}}} \nonumber \\
&=& \frac{\beta}{\log{2}} \tr{ L \rho_{\beta}  } + \log_{2}{Z_{\beta}},
\end{eqnarray}

where the trace in the first term can be written as the following summation
\begin{eqnarray}
\tr{{L} \rho_{\beta}}= \sum\limits_{i=1}^{N} \lambda_{i} \nu_{i}(\beta)=\sum\limits_{i=C+1}^{N} \lambda_{i} \frac{e^{-\beta \lambda_{i}}}{Z_{\beta}},
\end{eqnarray} 
the last step is justified by the fact that $\lambda_1,...,\lambda_C=0$ for a network with $C$ connected components. It is worth mentioning that the isolated nodes are considered to be separate components and are included in $C$.

A mean-field approximation of the above summation can be obtained by neglecting the higher-order terms as follows:
\begin{eqnarray}
\langle\lambda \nu(\beta)\rangle &=& \langle(\lambda -\bar{\lambda}+\bar{\lambda})(\nu(\beta)-\bar{\nu}(\beta)+\bar{\nu}(\beta))\rangle\nonumber\\
&=&\bar{\lambda}\bar{\nu}(\beta)+\langle(\lambda -\bar{\lambda})(\nu(\beta)-\bar{\nu}(\beta))\rangle \nonumber\\
&\approx&\bar{\lambda}\bar{\nu}(\beta).
\end{eqnarray}
To increase the precision, the terms in the summation corresponding to $\lambda_{i}=0$ must be excluded from the mean values of both sets of eigenvalues. Consequently, the mean-value for the Laplacian matrix follows
\begin{eqnarray}
\bar{\lambda}=\frac{1}{N-C}\sum_{i=C+1}^{N}\lambda_{i}=\frac{2m}{N-C},
\end{eqnarray}
and for the density matrix
\begin{eqnarray}
\bar{\nu}(\tau)=\frac{1}{N-C}\sum_{i=C+1}^{N}\frac{e^{-\tau \lambda_{i}}}{Z(\tau)} = \frac{1}{N-C}\frac{Z(\tau)-C}{Z(\tau)}.\nonumber
\end{eqnarray}
It follows that
\begin{eqnarray}
\tr{{L} \rho}&=&(N-C)\langle\lambda \nu(\tau)\rangle \nonumber \\
&\approx&\frac{2m}{N-C} \frac{Z_{\beta}-C}{Z_{\beta}}
\end{eqnarray}
which, for a network with no isolated nodes and only one connected component ($C=1$), and comparably large size $N \gg  1$, it reduces to 
\begin{eqnarray}\label{eq:2nd_term}
\tr{L \rho_{\beta}}\approx \bar{k}\frac{Z_{\beta}-1}{Z_{\beta}},
\end{eqnarray}
where $\frac{2m}{N-1}\approx \frac{2m}{N} = \bar{k} $ is the average degree of nodes. 

From here, it is straightforwad to combine Eq.~\ref{eq:entropy} and Eq.~\ref{eq:2nd_term} to obtain the mean-field entropy
\begin{eqnarray}
S^{MF}_{\beta} = \frac{1}{\log 2} ( \beta  \bar{k} \frac{Z_{\beta}-1}{Z_{\beta}} + \log Z_{\beta} ).
\end{eqnarray}

Whereas, for networks with isolated nodes and disconnected components the mean-field entropy reads:
\begin{eqnarray}
S^{MF}_{\beta} = \frac{1}{\log 2} ( \beta \frac{2m}{N-C} \frac{Z_{\beta}-C}{Z_{\beta}} + \log Z_{\beta} ). 
\end{eqnarray}

\textbf{Multiscale derivations.} For small scales the partition function can be written as $Z_{\beta} = \tr{e^{\beta {L}}} \approx \tr{I} - \beta \tr{{L}} = N - 2 \beta m $ and the density matrix follows $ \rho_{\beta} = \frac{1}{Z_{\beta}} e^{-\beta {L}} \approx \frac{1}{Z_{\beta}} ( I - \beta {L}  ) $. 

If the propagation time goes to zero limit $\beta \rightarrow 0 $, it can be shown that the density matrix is $\rho_{0}= I/N$ and the Von Neumann entropy depends, only, on the network size $S_{0} = - \sum\limits_{i=1}^{N} 1/N \log_{2} (1/N) = \log_{2} (N)$. 

Assume the size of original network $G$ is $N$. Then the size of perturbed network after removal of a node $G'_{x}$ (See Fig.~\ref{fig:disintegration}), is $N-1$ and the size of the star network corresponding to the detached node $\delta G_{x}$ depends on its degree $k_{x} + 1$. Therefore, the entanglement at $\beta \rightarrow 0$ 
\begin{eqnarray}
M_{0}(x) &=& [ \log_{2}(N-1) + \log_{2}(k_{x}+1)   ] - \log_{2}(N) \nonumber \\
&=& \log_{2}(\frac{N-1}{N}) + \log_{2}(k_{x}+1) \nonumber \\
&\approx & \log_{2}(k_{x}+1)
\end{eqnarray}
is proportional to the degree of the removed node. This proves that entanglement centrality and degree centrality coincide, for very small $\beta$.

Note that, for a network with $C$ connected components, the Laplacian matrix has exactly $C$ zero eigenvalues, while all other eigenvalues are greater than zero. Therefore, the partition function can, generally, be rewritten as $Z_{\beta} =  C + \sum\limits_{i=C+1}^{N} e^{-\beta \lambda_{i}} $ and approximated as $Z_{\beta} \approx C $, for large $\beta$. Also, Taylor expanding the logarithm of partition function around this point, one can find $ \log Z_{\beta} \approx \frac{Z_{\beta} - C }{Z_{\beta}} $. We put this result into Eq.~\ref{eq:MF_precise} to find the mean-field entropy at large $\beta$:
\begin{eqnarray}
S^{MF}_{\beta} \approx  ( \beta \frac{2m}{N-C} + 1 ) \log_{2} Z_{\beta},
\end{eqnarray}
which, in case of $N \gg  C$ becomes $(\beta \bar{k}+1) \log_{2} Z_{\beta}$ which can be approximated as
\begin{eqnarray}
S^{MF}_{\beta} \approx \beta \bar{k} \log_{2} Z_{\beta},
\end{eqnarray}
since $\beta \bar{k} \gg 1$. Also, in the limit case the above equation becomes
\begin{eqnarray}
\lim_{\beta \rightarrow \infty} S^{MF}_{\beta} \approx \beta \bar{k} \log_{2} C.
\end{eqnarray}

The star network corresponding to the removed node has only one connected component $C_{x}=1$. As $\log_{2}{1}=0$, the entropy follows $S^{MF}_{\infty}(\delta G_{x}) = 0$ for the star network. Let the number of connected components in $G$ and $G'$ be, respectively $C$ and $C'$, and their average numbers indicated by $\bar{k}$ and $\bar{k'}$. The entanglement, at the limit of large $\beta \rightarrow \infty$ follows
\begin{eqnarray}
M_{\beta}(x) =  \beta (\bar{k'} \log_{2}C' - \bar{k} \log_{2}C ).
\end{eqnarray}
In case the network is large, the removal of one node does not change its average degree dramatically $\bar{k} \approx \bar{k'}$. Therefore, the entanglement can be reduced to 
\begin{eqnarray}
M_{\beta}(x) =  \beta \bar{k} \log_{2}{ \frac{C'_{x}}{C}}.
\end{eqnarray}
Of course, in case the initial network is completely connected ($C=1$), we obtain 
\begin{eqnarray}
M_{\beta}(x) =  \beta \bar{k} \log_{2}{C'_{x}},
\end{eqnarray}
which is the case for all the synthetic networks considered in this work.

Finally, using Eq.\ref{eq:MF_approx}, one can write the entanglement of node $x$ as
\begin{eqnarray}
M_{\beta}(x) \approx (\beta \bar{k} + 1 ) \log_{2}{\frac{Z'_{\beta}(x)}{Z_{\beta}}}
\end{eqnarray}
at the meso-scale. Consequently, the collective entanglement (See Fig.~\ref{fig:disintegration}), follows
\begin{eqnarray}
\bar{M}_{\beta} \approx \frac{\beta \bar{k} + 1 }{N} \sum\limits_{x=1}^{N} \log_{2}{\frac{Z'_{\beta}(x)}{Z_{\beta}}}.
\end{eqnarray}
Taylor expanding each term in the summation around its minimum ($Z'_{\beta}(x)=Z_{\beta}$) and keeping only the first order term, we obtain
\begin{eqnarray}
\bar{M}_{\beta} &\approx& \frac{\beta \bar{k} + 1 }{N \log{2}} \sum\limits_{x=1}^{N} [ 0   + \frac{Z'_{\beta}(x) - Z_{\beta} }{Z_{\beta}}    ] \nonumber\\
&=&  \frac{\beta \bar{k} + 1 }{N Z_{\beta} \log{2}} \sum\limits_{x=1}^{N}\Delta Z_{\beta}(x),
\end{eqnarray}

where $\Delta Z_{\beta}(x) = Z'_{\beta}(x) - Z_{\beta}$. As $M_{\beta}$ nears its minimum, higher precision of the above linearization is expected. The scale at which the collective entanglement is at its minimum defines $\beta_{c}$. Finally, the entanglement centrality of node $x$ at $\beta_{c}$ follows
\begin{eqnarray}
M_{\beta_{c}}(x) = \frac{\beta \bar{k} + 1 }{N Z_{\beta} \log{2}} \Delta Z_{\beta_{c}}(x)
\end{eqnarray}

\textbf{Centrality measures.} A variety of centrality measures have been adopted, in the literature, to find the relative importance of the nodes for network integrity. In this section, we, briefly, review some of them that are used through this paper, including degree, betweenness, closeness, eigenvector, PageRank and clustering centrality.

\emph{Degree Centrality}. In an undirected network, the degree of each node is the number of its connections. Consequently, the degree centrality considers a node with higher number of connections more influential and, therefore, more important. Let $A$ be the adjacency matrix, where $A_{ij}=1$ encodes a connection between nodes $i$ and $j$ while $A_{ij}=0$ shows that they are not connected. Thus, the degree $k_{i}$ of node $i$ is given by $\sum\limits_{j=1}^{N}A_{ij}=k_{i}$.

\emph{Closeness Centrality}. It measures the importance of the node based on its average distance from the others, determined by the shortest path length. The shortest path between two nodes $i$ and $j$ is a path --i.e. sequence of links---connecting them that has minimum number of links. Let the average length of shortest paths connecting node $i$ to all the nodes of the network be $g_{i}$, the closeness centrality of node $i$ is given by $c_{i}=1/g_{i}$, indicating how close the node is to other nodes on average.

\emph{Betweenness Centrality}. According to betweenness centrality, a node's importance is determined by the number of shortest
paths that pass through it, connecting other nodes. In other words, assuming the shortest path to be the dominant pathway of information flow between the nodes, a node with high betweenness centrality is fundamental for node-node communications.

\emph{Eigenvector Centrality}. This centrality measure assesses the importance of a node, by the importance of their neighbors. Let $e_{i}$ be the eigenvector centrality of node $i$ which depends on the sum of eigenvector centrality of its neighbors followed by $e_{i}=\frac{1}{\alpha}\sum\limits_{i=1}^{N}A_{ij}e_{j}$, where $\alpha$ is a constant. Interestingly, it leads to an eigenvalue problem $ \mathbf{A}\vec{e} = \alpha \vec{e} $, for which the largest eigenvalue is considered to ensure the positivity of the components of the eigenvector.

\emph{Page Rank Centrality}.  Originally, this measure has been designed to investigate the world wide web. It is based on the definition of non-absorbed random walks, governed by the google matrix, on top of networks. According to this measure, the centrality of each node is proportional to the probability that the random walker visits it.\cite{brin1998}

\emph{Clustering Centrality}. Clustering centrality is based upon the definition of local clustering coefficient of nodes, which measures how densely the neighboring nodes are connected\cite{Watts1998}. More specifically, clustering coefficient of each node is proportional to the number of triads it shapes with the other nodes.

\heading{Contributions} AG performed the theoretical analysis, the numerical experiments and wrote the paper. MS performed the numerical experiments and the data analysis. JB performed part of the theoretical analysis and wrote the manuscript. MDD conceived and designed the study and wrote the manuscript. 

\heading{Competing financial interests} The authors declare no competing financial interests.

\bibliographystyle{naturemag}

\begin{small}
\bibliography{biblio}
%\begin{thebibliography}{10}
%\end{thebibliography}
\end{small}

%%%%%%%%%%%%%%%%%%%%%%%%%%%%%%%%%%%%

%\begin{figure}[!h]
%\centering
%\includegraphics[width=12cm]{Fig1.pdf}
%\caption{\label{fig:label}\small{\textbf{Short title.} Caption.}}
%\end{figure}

\end{document}